# Distribution of proper motions in spherical star clusters

M. Wybo and H. Dejonghe

Universiteit Gent, Sterrenkundig Observatorium, Krijgslaan 281, B–9000 Gent, Belgium

**Abstract.** Along with data on radial velocities, more and more data on proper motions of individual stars of globular clusters are becoming available.

Their usage was until now rather limited. It was mostly restricted to determining cluster membership of individual stars and to determining the spatial velocity of the cluster. We will study the two dimensional distribution of the proper motions, in order to clarify the relation between the projection of the velocities and the dynamical structure. Obviously some assumptions on the dynamics of the system have to be made. In this study we chose a Plummer model. This complements an earlier study about line profiles.

The three velocity dispersions in the main directions are determined and compared with the three spherical velocity dispersions. Two parameters analogous to Binney's anisotropic $\beta$-parameter are defined. Moreover, since the proper motion components are distance dependent and the line of sight velocity dispersion is distance independent, the distance can be determined.

## 1. Introduction

One of the greatest challenges in astrometry is the accurate measurement of minute relative variations in the positions of objects in the sky, in order to measure proper motions. Recent advances in data acquisition techniques, largely symbolized by the Hipparcos Mission, will produce a large quantity of proper motion data of high quality. The more traditional approach simply takes advantage of the lapse of time: the longer the time span that is covered, the better proper motions are determined, at least if the first-epoch medium is sufficiently well behaved. No wonder that useful proper motions are also becoming available in those areas of the sky that happened to be of particular interest to astronomers (then and now). Globular clusters are such areas, and plates exist that are almost a century old. There is already a body of data on proper motions for globular clusters (Cudworth & Smetanka 1992, Rees 1992 and references therein). Similarly, other data sets, located in popular fields in our own Galaxy and the Galactic Bulge, are being produced (Minniti 1992).

In the case of globulars, one of the uses of proper motions consists in the determination of cluster membership by eliminating stars with anomalously large proper motions. From the astrophysical point of view, such a procedure produces cleaner color-magnitude diagrams. Another more recent use is the determination of the proper motion of the globular itself (Cudworth and Hanson 1993). On the other hand, proper motions are useful for what they directly mean: they provide kinematical information. They are similar to the line-of-sight velocity, which has proven to be very useful in galactic cluster and galaxy research. Therefore we set out to study the theoretical aspects involved in calculating and interpreting proper motions. This we do very much in the same way as was done by one of us for the line-of-sight velocities (Dejonghe 1987, hereafter Paper I).

In sections 2 and 3, we consider a one-parameter sequence of Plummer models, which covers a variety in orbital structure. It is well-known that the Plummer model is not fit to describe stellar systems, be they globulars or spherical galaxies. Nevertheless, we consider the Plummer model because of its analytical simplicity, and we are confident that our results are applicable to real stellar systems, at least in a qualitative sense. A similar statement can, in hindsight, be made about the models in Paper I.

We study the projected proper motions and compare our results with the earlier calculated line profiles for each member of the Plummer family. Conversely, we show what dynamical information can be extracted from observed proper motions. In particular, it was our goal to get information on the orbital structure (i.e. radial or tangential orbits) of the system, *regardless of our distance to the system*.

Since, as was already mentioned, the distance-dependent proper motions are very similar in nature to the distance-independent line-of-sight velocities, it is clear that the simultaneous knowledge of line-of-sight velocity distributions and proper motion distributions will constrain the distance to the system (section 4). A first attempt at this can be found in Lupton, Gunn & Griffin (1987).

Much of the work can be carried out analytically, which is not only a useful check on the numerical calculations, but gives this work, we hope, an additional quality which is increasingly rarely found.

## 2. The theoretical framework

### 2.1. The Plummer family

We consider anisotropic distribution functions $F(E, L)$, which are functions of the specific binding energy $E = \psi(r) - \frac{1}{2}(v_r^2 + v_\vartheta^2 + v_\varphi^2)$, whereby $\psi(r) = -V(r) \geq 0$ is the potential function the angular momentum vector.

The Plummer model has the well known potential

$$\psi(r) = (1+r^2)^{-\frac{1}{2}}, \qquad (1)$$

expressed in some suitable units, with the corresponding mass density

$$\rho(r) = \frac{3}{4\pi}(1+r^2)^{-\frac{5}{2}}. \qquad (2)$$

Dynamical anisotropic models for the Plummer mass density can be constructed considering the augmented mass densities

$$\tilde{\rho}(\psi, r) = \frac{3}{4\pi}\psi^{5-q}(1+r^2)^{-\frac{q}{2}}, \qquad (3)$$

dependent on the parameter $q$, but yielding the same physical mass density for all $q$, which is obtained by substituting Eq.(1) in Eq.(3). The distribution function $F(E, L)$ that corresponds to a $\tilde{\rho}(\psi, r)$ turns out to be (Dejonghe, 1986)

$$F(E, L) = \frac{3\,\Gamma(6-q)}{2\,(2\pi)^{\frac{5}{2}}\Gamma(\frac{9}{2}-q)} E^{\frac{7}{2}-q} \;_2F_1\left(\frac{q}{2}, q-\frac{7}{2}; 1; \frac{L^2}{2E}\right)$$
$$L^2 < 2E \qquad (4)$$

$$F(E, L) = \frac{3\,\Gamma(6-q)}{2\,(2\pi)^{\frac{5}{2}}\Gamma(\frac{9}{2}-\frac{q}{2})\Gamma(1-\frac{q}{2})} \left(\frac{2E}{L^2}\right)^{\frac{q}{2}} E^{\frac{7}{2}-q}$$
$$\times \;_2F_1\left(\frac{q}{2}, \frac{q}{2}; \frac{9}{2}-\frac{q}{2}; \frac{2E}{L^2}\right) \qquad L^2 > 2E \qquad (5)$$

where $\Gamma(x)$ denotes the gamma function. Positiveness of $F(E, L)$ restricts the value of $q$ to $q \leq 2$. The physical meaning of $q$ is most obvious when considering Binney's anisotropy parameter (Binney & Tremaine, 1987, p204)

$$\beta = 1 - \frac{\sigma_\varphi^2}{\sigma_r^2} = 1 - \frac{\sigma_\vartheta^2}{\sigma_r^2} = \frac{q}{2}\frac{r^2}{1+r^2}. \qquad (6)$$

The important point to note is that $q$ and $\beta$ have the same sign. Hence radial orbits are prevalent if $\beta > 0$, and more tangential orbits are prevalent in a $\beta < 0$ cluster. If $\beta = 0$ then both types occur in equal amounts in an isotropic sense. If $q$ becomes very negative, then $\sigma_r^2 \to 0$, so in the limit $q \to -\infty$ the stellar system only contains circular orbits. The other limiting case $q = 2$ has a $\beta$ which tends to 1 for great distances of $r$, so in the outer regions only radial orbits are present. We classify the models into radial systems, with $0 < q \leq 2$, tangential systems, with $q < 0$, and isotropic systems, with $q = \beta = 0$. The latter case is normally called a Plummer model.

2.2. Notations - normalisations

The velocity $\mathbf{v}$ of a star in spherical coordinates is in obvious notations: $\mathbf{v} = v_r \mathbf{e}_r + v_\varphi \mathbf{e}_\varphi + v_\vartheta \mathbf{e}_\vartheta$. Generally we cannot observe these components: at best we obtain the projection of $\mathbf{v}$ along the line of sight which we denote by $v_p$ and the proper motion which is the projected velocity on the plane of the sky. This proper motion is normally measured in arcsec/year in some coordinate system, say $(\mu_\alpha, \mu_\delta)$. For this paper it is useful to define a system of perpendicular axes so that the proper motion is decomposed in a component $\mu_c$ which is measured along the projected radius, and $\mu_t$ which is measured perpendicular to it. The velocity components $v_c$ and $v_t$ are obtained by the relations $v_c = \mu_c \Delta$ and $v_t = \mu_t \Delta$, whereby $\Delta$ is the distance to the system. In the following, we will refer to $v_c$ as the center velocity (and not radial velocity, since otherwise we would add to the confusion about the term radial) and $v_t$ as the transverse velocity. So in this system the velocity $\mathbf{v}$ is: $\mathbf{v} = v_c \mathbf{e}_c + v_t \mathbf{e}_t + v_p \mathbf{e}_p$. We will transform the frame $(\mathbf{e}_r, \mathbf{e}_\varphi, \mathbf{e}_\vartheta)$ to $(\mathbf{e}_c, \mathbf{e}_t, \mathbf{e}_p)$, considering the isotropy in $\vartheta$ and $\varphi$. The axes $\mathbf{e}_r, \mathbf{e}_c$ and $\mathbf{e}_p$ form a plane $\alpha$. We rotate the axes $\mathbf{e}_\vartheta$ and $\mathbf{e}_\varphi$ so that $\mathbf{e}_\varphi$ lies in $\alpha$. Then $\mathbf{e}_t$ is lying according to $\mathbf{e}_\vartheta$, since both vectors are perpendicular to $\alpha$. Let $r$ be the distance and $r_p$ be the projected distance from the center of the cluster to the star; and let $z$ be the algebraic distance from the star to the plane of the sky which cuts through the center of the system. Figure 1 gives a perspective view of both coordinate frames and the distances $r$, $r_p$ and $z$. The transformation is a simple rotation around $\mathbf{e}_t = \mathbf{e}_\vartheta$ and the formulae are then:

$$\begin{cases} v_r = \frac{r_p}{r}v_c - \frac{z}{r}v_p \\ v_\varphi = \frac{z}{r}v_c + \frac{r_p}{r}v_p \\ v_\vartheta = v_t \end{cases} \Leftrightarrow \begin{cases} v_c = \frac{r_p}{r}v_r + \frac{z}{r}v_\varphi \\ v_p = -\frac{z}{r}v_r + \frac{r_p}{r}v_\varphi \\ v_t = v_\vartheta \end{cases} \qquad (7)$$

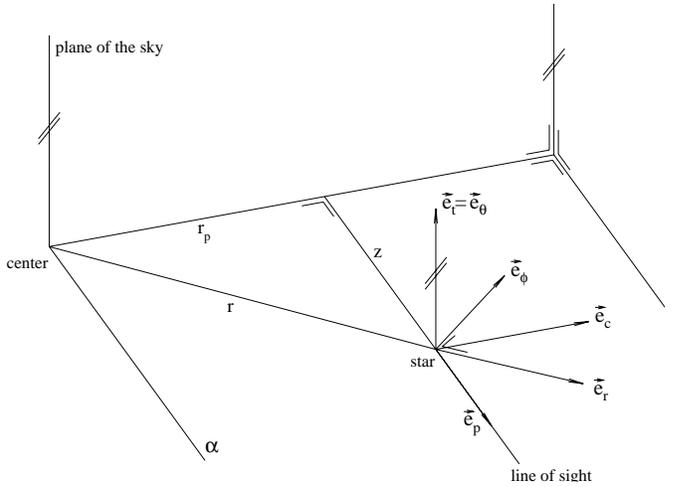

**Fig. 1.** The geometrical relations between the frame related to the spherical coordinate system $(\mathbf{e}_r, \mathbf{e}_\varphi, \mathbf{e}_\vartheta)$ and the frame related to the plane of sky and the line of sight $(\mathbf{e}_c, \mathbf{e}_t, \mathbf{e}_p)$.

The projection $f_p(r_p)$ of any scalar quantity $f(r)$ on the plane of the sky is

$$f_p(r_p) = \int_{-\infty}^{+\infty} f(r)\,dz = 2\int_{r_p}^{+\infty} f(r)\,\frac{r\,dr}{\sqrt{r^2 - r_p^2}}. \qquad (8)$$

For example, the projected mass density in the Plummer model reads:

$$\rho_p(r_p) = \frac{1}{\pi}(1+r_p^2)^{-2} = \frac{\psi_p^4}{\pi}, \qquad (9)$$

whereby $\psi_p \equiv \psi(r_p) = (1+r_p^2)^{-\frac{1}{2}}$ is the potential on the projected distance $r_p$ from the center.

Hence, the cumulative mass as a function of $r_p$ is

$$M_p(r_p) = 2\pi \int_0^{r_p} \rho_p(r_p)\, r_p\, dr_p = \frac{r_p^2}{1 + r_p^2}, \qquad (10)$$

which can be used to define the core radius as the radius at which $\rho_p(r_p)$ drops to half of its central value: $r_c^2 = \sqrt{2} - 1 \to r_c \simeq 0.64$.

To determine the proper motion distribution, it is convenient to redefine the independent variables (the projected velocities): at every radius, the velocity is normalized to unity for the largest proper motion corresponding to bound orbits. The maximum velocity which occurs at the distance $r_p$ from the center is given by $v_{\max} = \sqrt{2\psi_p}$. The normalized velocity components $\chi, \eta, \zeta$ are thus

$$\chi = \frac{v_c}{\sqrt{2\,\psi_p}}, \quad \eta = \frac{v_t}{\sqrt{2\,\psi_p}}, \quad \zeta = \frac{v_p}{\sqrt{2\,\psi_p}}, \qquad (11)$$

and we also denote the normalized modulus of the proper motion by

$$v = \sqrt{\chi^2 + \eta^2}. \qquad (12)$$

Note also that this normalization dispenses with the need to distinguish between the proper motion ($\mu_c, \mu_t$) and the proper motion velocities ($v_c, v_t$). The energy and angular momentum expressed in the normalized velocities are:

$$2E = 2\psi - (v_c^2 + v_t^2 + v_p^2) = 2\psi_p\left(\frac{\psi}{\psi_p} - (\chi^2 + \eta^2 + \zeta^2)\right) \qquad (13)$$

$$L^2 = r^2(v_\varphi^2 + v_\vartheta^2) = 2\,\psi_p\left(r^2\,\eta^2 + (z\,\chi + r_p\,\zeta)^2\right). \qquad (14)$$

### 3. Proper motion distributions

Let $\mathrm{pm}_{r_p}(v_c, v_t)$ be the proper motion distribution (PM-distribution) at $r_p$. To calculate this PM-distribution we need to perform a double integration. The first integral averages over the velocity-component along the line of sight ($v_p$) and the second integrates through the cluster ($z$). The PM-distribution is given by:

$$\mathrm{pm}_{r_p}(v_c, v_t)\, dv_c\, dv_t$$
$$= dv_c\, dv_t \int_{r_p}^{r_{\max}} \frac{dr^2}{\sqrt{r^2 - r_p^2}} \int F(E, L)\, dv_p \qquad (15)$$

and depends of course on the distribution function $F(E, L)$. The boundary $r_{\max}$ is the value of $r$ for which the escape velocity is reached, or: $1 + r_{\max}^2 = \psi_{\max}^{-2} = 1/(\psi_p^2\, v^4)$.

It is useful to normalize the proper motion probability distribution to one, though it really is proportional to the projected mass density given in Eq. (9).

*3.1. Proper motion in the case $q = -2n$ ($n$ a natural number)*

In this case the PM-distribution can be calculated exactly. The hypergeometric series in the distribution function $F(E, L)$ is then given by

$$F(E, L)$$
$$= \frac{3\Gamma(6 + 2n)}{2(2\pi)^{\frac{5}{2}}\Gamma(\frac{9}{2} + 2n)} E^{\frac{7}{2}+2n}\, {}_2F_1(-n, -2n - \frac{7}{2}; 1; \frac{L^2}{2E}) \qquad (16)$$

and terminates. Moreover, in this case the distribution function need not be specified by the double prescription (4) and (5). The PM-distribution (15) becomes after normalisation:

$$\mathrm{pm}_{r_p}(\chi, \eta)$$
$$= \frac{2\pi}{\psi_p^3} \int_{r_p}^{r_{\max}} \frac{dr^2}{\sqrt{r^2 - r_p^2}} \int_{-\zeta_{\max}}^{\zeta_{\max}} F(E, L)\,\sqrt{2\psi_p}\, d\zeta \qquad (17)$$

where $\zeta_{\max}$ is the value of $\zeta$ for which $E$ equals zero, or $\zeta_{\max} = ((\psi/\psi_p) - v^2)^{1/2}$. The integration over the velocity component $\zeta$ results in

$$\int_{-\zeta_{\max}}^{\zeta_{\max}} F(E, L)\sqrt{2\psi_p}\, d\zeta$$
$$= \frac{3\,\Gamma(6+2n)\,\psi_p^{2n+1}}{4\pi} \sum_{m=0}^n \binom{n}{m} \sum_{i=0}^m \frac{\Gamma(i+\frac{1}{2})}{\Gamma(m-i+1)}(r\eta)^{2(m-i)}$$
$$\times \sum_{j=0}^i \frac{r_p^{2j}\,(z\chi)^{2(i-j)}\,(\zeta_{\max}^2)^{j+2n-m+4}}{\Gamma(j+1)\Gamma(i-j+\frac{1}{2})\Gamma(2n-m+j+5)\Gamma(i-j+1)}. \qquad (18)$$

To avoid cumbersome notation, we set $h = 2n - m + j + 4$. For the integration through the cluster we need to calculate

$$I_c = \int_{r_p}^{r_{\max}} \frac{dr^2}{\sqrt{r^2 - r_p^2}} \left(\frac{\psi}{\psi_p} - v^2\right)^h (r^2)^{m-i}\, (z^2)^{i-j}. \qquad (19)$$

With the substitution $x = \psi_p(1 + r^2)^{\frac{1}{2}}$ this leads to

$$I_c = \psi_p^{2j-2m-1}$$
$$\times \int_1^{1/v^2} (x^2 - 1)^{i-j+\frac{1}{2}} \left(\frac{1}{x} - v^2\right)^h (x^2 - \psi_p^2)^{m-i}\, dx^2. \qquad (20)$$

The powers of the last two factors are positive integers so they can be expanded. Upon interchanging the integration and the summation signs, and using Euler's integral the above equation transforms into:

$$I_c = \psi_p^{2j-2m-1}\, \frac{\Gamma(i-j+\frac{1}{2})}{\Gamma(i-j+\frac{3}{2})} \left(\frac{1}{v^4} - 1\right)^{i-j+\frac{1}{2}}$$
$$\times \sum_{k=0}^h \binom{h}{k}(-v^2)^{h-k} \sum_{l=0}^{m-i} \binom{m-i}{l}(-\psi_p^2)^{m-i-l}$$
$$\times {}_2F_1(\frac{k}{2}-l,\, i-j+\frac{1}{2};\, i-j+\frac{3}{2};\, 1-\frac{1}{v^4}). \qquad (21)$$

The PM-distribution (17), using polar coordinates namely $\chi = v\cos\alpha$ and $\eta = v\sin\alpha$ and a transformation of the hypergeometric function then becomes:

$$\mathrm{pm}_{r_p}(\chi, \eta)$$
$$= \frac{3}{4\pi}\Gamma(6 + 2n)\psi_p^{2n}\sqrt{1 - v^4}\, v^6$$
$$\times \sum_{m=0}^n \binom{n}{m} \sum_{i=0}^m (-1)^i\, \Gamma(i+\frac{1}{2})(\sin^2\alpha)^{m-i}$$
$$\times \sum_{j=0}^i \frac{(-1)^j\, r_p^{2j}}{\Gamma(j+1)\,\Gamma(i-j+1)\,\Gamma(i-j+\frac{3}{2})}\left(\cos^2\alpha\,\frac{1-v^4}{\psi_p^2}\right)^{i-j}$$

$$\times \sum_{l=0} \frac{(-1)^{\ldots}}{\psi_p^{2l}} \frac{(v^{\ldots})^{\ldots}}{\Gamma(l+1)\Gamma(m-i-l+1)}$$

$$\times \sum_{k=0}^{h} \frac{(-1)^k}{\Gamma(k+1)\Gamma(h-k+1)} \, _2F_1(\frac{k}{2}-l, 1; i-j+\frac{3}{2}; 1-v^4). \quad (22)$$

This result can be made considerably simpler for zero velocity. In this case integration over the velocity component $\zeta$ as given in Eq. (18) yields:

$$\int_{-\zeta_{\max}}^{\zeta_{\max}} F(E,L)\sqrt{2\psi_p}\, d\zeta$$
$$= \frac{3\,(5+2n)}{4\,\pi} \psi_p^{2n+1} (\zeta_{\max})^{2n+4} \, _2F_1(-n, \frac{1}{2}; 1; -r_p^2), \quad (23)$$

since all positive powers of $\eta$ and $\chi$ vanish. The integration through the cluster results in a beta-function, so that the PM-distribution for zero velocity is given by:

$$\langle \mathrm{pm}_{r_p}(0,0)\rangle_{q=-2n}$$
$$= \frac{3\,(5+2n)}{4\sqrt{\pi}} \psi_p^{2n} \frac{\Gamma(n+\frac{3}{2})}{\Gamma(n+2)} \, _2F_1(-n, \frac{1}{2}; 1; -r_p^2). \quad (24)$$

This expression can also be obtained from Eq.(22) by taking the limit for $v \to 0$.

Equation (22) can also be expressed in elementary functions for specific $q$, using Gradshteyn and Ryzhik (1965, Eqs. 9.121). The simplest case is the isotropic one ($q=0$), which reads:

$$\mathrm{pm}_{r_p}(\chi, \eta)$$
$$= \frac{15}{4\pi} \left[ v^2(2v^4-7)\sqrt{1-v^4} + (1+12\,v^2) \arccos v^2 \right]$$
$$+ \frac{30}{\pi} v^6 \ln \frac{v^2}{1+\sqrt{1-v^4}}. \quad (25)$$

If $v \to 1$, then $\mathrm{pm}_{r_p}(\chi,\eta)$ goes to the expected value zero. If $v \to 0$, then $\mathrm{pm}_{r_p}(0,0) = 15/8$. Equation (25) shows that in this case the PM-distribution does only depend on the velocity $v^2$, which is not surprising. More surprising however is that all normalized PM-distributions are the same irrespective of the projected distance $r_p$.

This result holds also for other distribution functions within the same potential, but which, of course, do not produce a self-consistent mass density. For example if $F(E) = A(2E)^{\frac{1}{2}+m}$, with $A$ some constant, the integration $I_c$ through the cluster and over the velocity component $\zeta$ (we work again with normalised velocities) results in:

$$\rho_p(r_p)\,\mathrm{pm}(\chi,\eta)$$
$$= 4A\,(2\,\psi_p)^{m+1} B(m+\frac{3}{2},\frac{1}{2})(-1)^{m+1}(v^2)^m \sqrt{1-v^4}$$
$$\times \sum_{i=0}^{m+1} \binom{m+1}{i}(-1)^i \, _2F_1(\frac{i}{2}, 1; \frac{3}{2}; 1-v^4). \quad (26)$$

The spatial mass density is given by:

$$\rho(r) = 4\pi \int F(E)\,v^2\,dv = 4\pi A(2\psi)^{m+2} B(\frac{3}{2}, m+\frac{3}{2}), \quad (27)$$

yielding a projected mass density

$$\rho_p(r_p) = 8\pi A\,(2\,\psi_p)^{m+1} B(\frac{3}{2}, m+\frac{3}{2}) B(\frac{1}{2}, \frac{m}{2}+\frac{1}{2}). \quad (28)$$

One easily checks that for $m=3$, using the appropriate analytical expressions of the hypergeometric function (see Gradsteyn & Ryzhik, 1965), the original isotropic Plummer model is recovered.

Another particular case is $q=-2$. Obviously, this is a more elaborate example. The PM-distribution (22) expressed in elementary functions, leads after some calculations to:

$$\mathrm{pm}_{r_p}(\chi,\eta)$$
$$= \frac{21}{32\pi}\sqrt{1-v^4}\,\Big((104v^8-866v^4-93)v^2\psi_p^2$$
$$\quad - v^2(56v^8+866v^4-67) + 16(6v^8-47v^4-4)\chi^2 r_p^2\psi_p^2\Big)$$
$$+ \frac{63}{32\pi}\arccos v^2\,\Big(40v^8-140v^4+1$$
$$\quad + (360v^8+100v^4+1)\psi_p^2 + 80(4v^4+1)v^2\chi^2 r_p^2\psi_p^2\Big)$$
$$+ \frac{63}{2\pi}v^6 \ln\frac{1+\sqrt{1-v^4}}{v^2}$$
$$\quad \times \Big(4v^4+20 - 11\psi_p^2 v^4 - 10v^2\chi^2 r_p^2\psi_p^2\Big) \quad (29)$$

Again, if $v \to 1$, then $\mathrm{pm}_{r_p}(\chi,\eta)$ goes to the expected value zero. And if $v \to 0$, then $\mathrm{pm}_{r_p}(0,0) = 63(1+\psi_p^2)/64$. The result here depends on the projected distance and on the velocity components. Later we will use this exact distribution to test our numerical results.

This kind of calculations could not be performed with the line profiles (see paper I), since for $q=-2n$ the power of $\zeta_{\max}^2$ in Eq. (18) is not an integer so this expression cannot be expanded.

### 3.2. Proper motion in the general anisotropic case

The distribution function $F(E,L)$ is given by Eqs. (4) and (5). The integral $\int F(E,L)\,d\zeta$ cannot be calculated in a general manner due to the cumbersome expression of $F(E,L)$ as a function of $\zeta$, so we have to work with approximations. However, we can calculate exactly

1. the *"top value"*: this is the probability of finding a star at $r_p$ without proper motion, i.e. $\mathrm{pm}_{r_p}(0,0)$, and this is found to be (see appendix A):
$$\mathrm{pm}_{r_p}(0,0) = \frac{3\,(5-q)\,\Gamma(\frac{3}{2}-\frac{q}{2})}{4\sqrt{\pi}\,\Gamma(2-\frac{q}{2})} \, _2F_1(\frac{q}{2},\frac{1}{2};1;\frac{r_p^2}{1+r_p^2}). \quad (30)$$
If $q=-2n$ this equation reduces, after a transformation of the hypergeometric function, to Eq.(24).

2. *all the moments* of the two-dimensional PM-distribution. They are determined in appendix B and are:
$$\langle v_c\,v_t\rangle^{2p,2s} = \frac{2^{p+s}}{\pi}(1+r_p^2)^{-\frac{p+s}{2}-2}\langle\chi\,\eta\rangle^{2p,2s}, \quad (31)$$
whereby the normalized moments are given by
$$\langle\chi\,\eta\rangle^{2p,2s} = \frac{3}{4\sqrt{\pi}}\Gamma\binom{p+\frac{1}{2},\,s+\frac{1}{2},\,6-q}{p+s+6-q}$$
$$\times \sum_{i=0}^{s}\binom{s}{i} x^{2i}\frac{(-1)^i\Gamma(\frac{1}{2}(p+s)+i+2)\left(\frac{q}{2}\right)_i}{\Gamma(i+1)\Gamma(\frac{1}{2}(p+s+5)+i)}$$
$$\times \, _3F_2\binom{\frac{q}{2}+i,\,\frac{1}{2},\,-(p+s-i);}{i+1,\,\frac{1}{2}(p+s+5)+i;}\,1\Bigg). \quad (32)$$

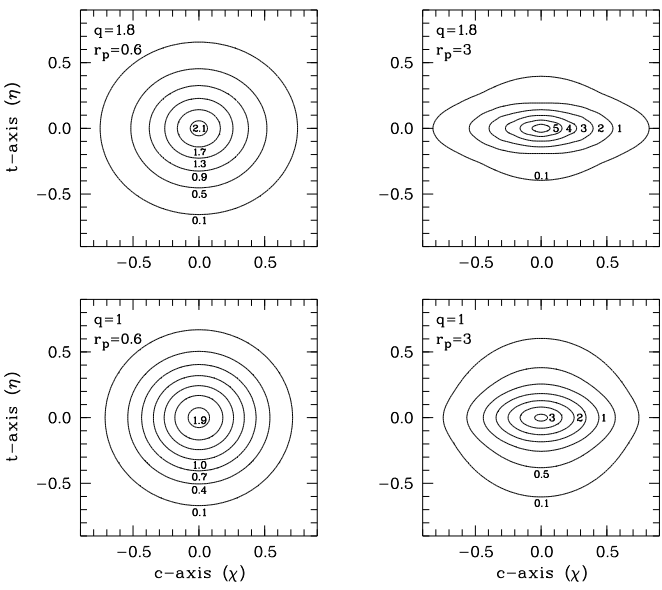
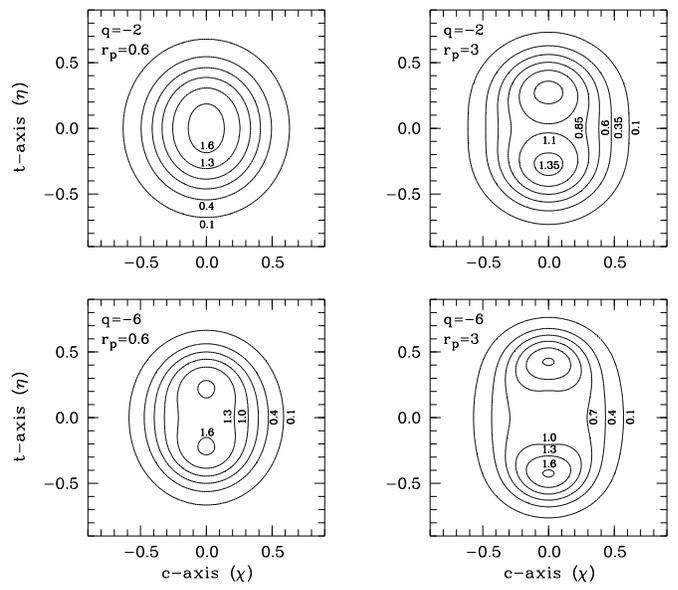

**Fig. 2.** Probability contours for the PM-distribution of different radial $q$-clusters. Left panels at the core radius ($r_p = 0.6$), right panels at five times the core radius.

**Fig. 3.** Probability contours for the PM-distribution of different tangential $q$-clusters. Left panels at the core radius ($r_p = 0.6$), right panels at five times the core radius.

In particular, two of them have special significance in view of later applications:

$$\langle \chi\, \eta \rangle^{2,0} = \frac{3\pi(12 - q)}{2^8(6 - q)} \qquad (33)$$

and

$$\langle \chi\, \eta \rangle^{0,2} = \frac{3\pi}{2^8(6 - q)} \left( 12 - 6q + \frac{5q}{1 + r_p^2} \right). \qquad (34)$$

It is a logical step to try to use the above results as an aid to calculate the PM-distributions. To that end, we conjecture that the distribution can be well approximated by the series

$$\mathrm{pm}_{r_p}(\chi, \eta) = \sum_{i,j} a_{ij} \left( 1 - \chi^2 - \eta^2 \right)^{\alpha + i} \eta^{2j}. \qquad (35)$$

We can determine the $a_{ij}$ coefficients by matching the moments of the approximated distribution in Eq. (35) with the true moments and by equating $\mathrm{pm}_{r_p}(0,0)$ from Eq. (30) to $\sum_i a_{i0}$. Therefore it is useful to obtain an analytical estimate for the exponent $\alpha$.

For small values of $r_p$ the PM-distribution is approximated by (see appendix C):

$$\mathrm{pm}_{r_p}(\chi, \eta) = \frac{3\,\Gamma(6 - q)\,\psi_p^{-q}}{2\sqrt{2\pi}\,\Gamma(\frac{11}{2} - q)} \left( 1 - v^2 \right)^{\frac{9}{2} - q}, \qquad (36)$$

and for large values of $r_p$ this approximation is:

$$\mathrm{pm}_{r_p}(\chi, \eta) =$$
$$\frac{3(5 - q)\,\Gamma(5 - \frac{q}{2})}{2\sqrt{2\pi}\,\Gamma(\frac{11 - q}{2})\,\Gamma(1 - \frac{q}{2})} \left(\frac{r_p^2}{1 + r_p^2}\right)^{-\frac{q}{2}} \left(1 - v^2\right)^{\frac{9 - q}{2}} \left(\eta^2\right)^{-\frac{q}{2}}. \qquad (37)$$

Hence we adopt $\alpha = 9/2 - q$.

In Figs. 2-3 we plot the contour lines of the distribution for some values of $q$ and $r_p$. The calculations of the two dimensional PM-distribution were checked twice, once by numerical integration at some points and once by comparison of the results with the exactly calculated distributions for $q = 0$ (isotropic case) and for $q = -2$. The matches were excellent for tangential systems as long as the $q$ values were not too negative ($q < -6$) and the distance $r_p$ was not too large. In the other cases, the individual terms in the calculation of the moments (32) tend to become very large, causing subtractive cancellation, so that the numerical solution for the coefficients and the subsequent expression (35) is correspondingly unreliable. For radial systems the series approximation did not deliver such good results for large positive ($> 1$) $q$ values or large $r_p$ values. The reason is that the PM-distribution does not approach zero uniformly on circles. Hence the figures that did not give a good match due to accuracy errors were produced by straightforward (but fairly slow) numerical integration.

Admittedly, the surfaces in Figs. 2-3 are only exact for the Plummer models we consider here. It is not unreasonable to assume, however, that our results carry over, at least in a qualitative sense, to more realistic models.

We immediately notice the difference between the PM-distributions for both cluster types. For the radial systems (Fig.2) the PM-distribution is unimodal and becomes more and more elongated along the center-axis. This feature is more prominent with increasing $q$ and $r_p$. This can be explained by the fact that our radial clusters have predominantly radial orbits in their outer parts. There is no contamination of the (spatial) tangential velocities in the center direction. For tangential systems, (Fig.3) the PM-distribution becomes bimodal towards the transverse direction. This feature is again more prominent with increasing $|q|$ and $r_p$. Again, this can be explained by noting that our tangential clusters have predominantly tangential orbits in their outer parts. There is no contamination of the (spatial) radial velocities in the tangential direction.

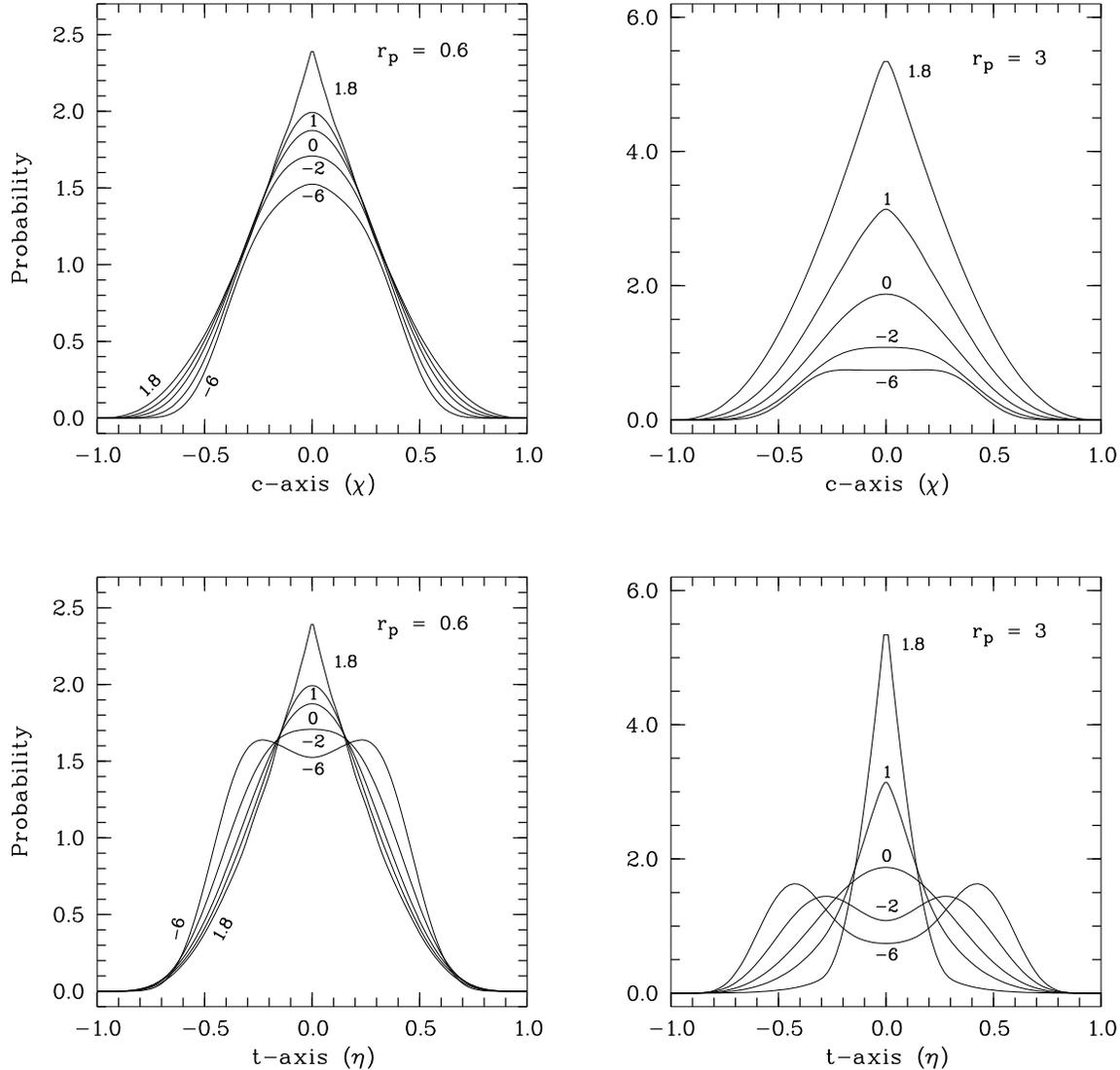

**Fig. 4.** These pictures show different cuts of the PM-distributions along both velocity axes, at the core radius ($r_p = 0.6$, left panels) and at five times the core radius (right panels). Values of $q$ are indicated in the figure.

It is important to realise that Figs. 2 and 3 provide an immediate way of determining qualitatively the orbital structure of star clusters! There is no need for a distance determination. In particular we note the striking difference between the right panels of Figs. 2 and 3, which show distribution at about 5 core radii, which is therefore a radius where a clearcut diagnosis should be possible if sufficient stars are present. A different pair is supposedly the lower-left panel of Fig. 2 and the upper-left one of 3 showing the PM-distribution at core radius of two clusters with not too different an orbital structure. Nevertheless, it does not seem hopeless to distinguish both with good quality data. Clearly the dispersions of the marginal distributions must play a key role, and we discuss them in the next section.

Finally, Fig. 4 shows, more quantitatively, the cuts of the distributions along both velocity axes. Here we see the dependence of the PM-distribution upon $\chi$ or $\eta$ for various $q$. The more positive $q$, the more peaked the PM-distribution is in the top. The more negative $q$, the more bimodal the distribution becomes. The effect is more pronounced at large projected radii.

## 4. The dispersions

The second order moments in the proper motions are of particular interest, since these are the most securely determined. The first order moments in these models are zero. Using Eq. (32), we find for the dispersions

$$\sigma_c^2(r_p) = \langle v_c^2 \rangle = \frac{2}{\sqrt{1+r_p^2}} \langle \chi \, \eta \rangle^{2,0} = \frac{3\pi(12-q)}{2^7(6-q)\sqrt{1+r_p^2}}, \quad (38)$$

and

$$\sigma_t^2(r_p) = \langle v_t^2 \rangle = \frac{2}{\sqrt{1+r_p^2}} \langle \chi \, \eta \rangle^{0,2}$$

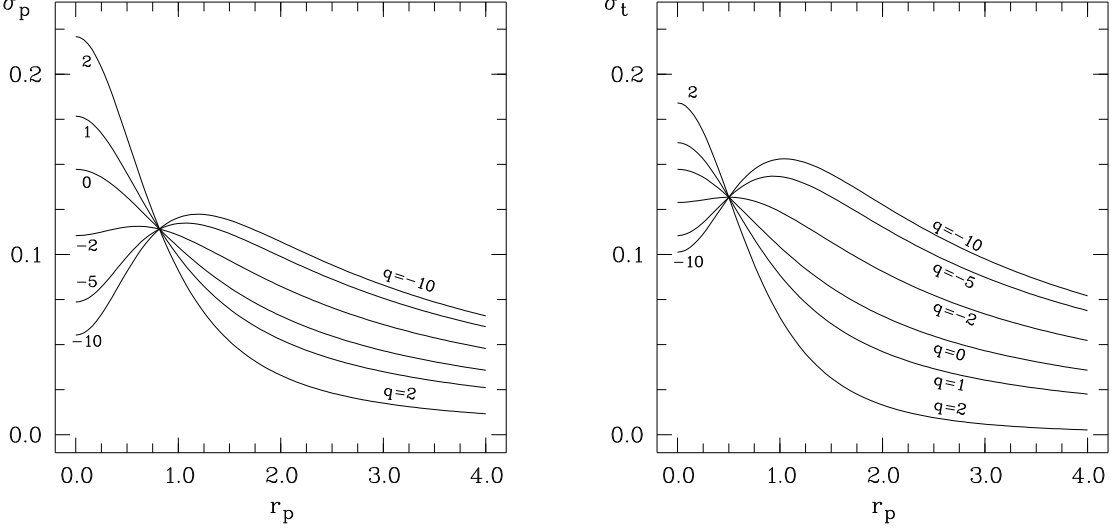

**Fig. 5.** The line of sight velocity dispersions and the transverse velocity dispersions are plotted for various $q$. Note the qualitative similarity.

$$= \frac{3\pi}{2^7(6-q)\sqrt{1+r_p^2}} \left(12 - 6q + \frac{5q}{1+r_p^2}\right). \quad (39)$$

This last result is very similar to the line of sight velocity dispersion calculated in paper I:

$$\sigma_p^2(r_p) = \frac{3\pi}{2^7(6-q)\sqrt{1+r_p^2}} \left(12 - 5q + \frac{5q}{1+r_p^2}\right). \quad (40)$$

In Fig.5 the line of sight velocity dispersion and the transverse dispersion are plotted against the radial distance $r_p$, for varying $q$. Both have qualitatively a very similar behaviour. In both panels of Fig.5 there are fixed points, due to the fact that $q$ enters linearly in both the numerator and the denominator of Eq.(39) and Eq.(40). For the line-of-sight velocity dispersions the fixed point is reached at $r_p = \sqrt{3/2}$ and has the functional value $\sigma_p^2 = (3\pi/64)\sqrt{3/5}$. In the transverse direction the fixed point is located at $r_p = 1/2$, with functional value $\sigma_t^2 = 3\pi/(32\sqrt{5})$. Also a maximum occurs for $\sigma_p^2$ at $r_p^2 = -(12+10q)/(12-5q)$ if $q \leq -6/5$ and for $\sigma_t^2$ at $r_p^2 = -(12+9q)/(12-6q)$ if $q \leq -4/3$. Both figures have in common that radial clusters look hotter at the center than tangential clusters, in both projections. On the contrary, at the edge the tangential clusters look hotter, as can be expected. The difference lies in the magnitude of the effect, which is greater for $\sigma_p^2$ than for $\sigma_t^2$ in the center, while the opposite is true at larger distances.

The center velocity dispersion on the contrary is a simple monotonous decreasing function of $r_p$. Moreover, for every $r_p$ it is an increasing function of the parameter $q$. So at each projected distance radial clusters give a higher contribution to the center velocity dispersion, which is quite obvious. At the center ($r_p = 0$), the proper motion velocity dispersions have the same value, namely $3\pi(12-q)/(2^7(6-q))$. The greater $q$, the higher the dispersions (first derivative is always positive). This feature was also noticed in paper I for the line of sight velocity dispersion and for the square of the transverse velocity.

With these values of the velocity dispersions, the total kinetic energy $T$ can easily be calculated. Let $\sigma_i^2(r_p)$ be the sum of the three projected velocity dispersions, so $\sigma_i^2(r_p) = \sigma_p^2(r_p) + \sigma_t^2(r_p) + \sigma_c^2(r_p)$, and $\sigma_s^2(r)$ be the sum of the three spherical velocity dispersions given by

$$\sigma_r^2(r) = \frac{1}{6-q}(1+r^2)^{-\frac{1}{2}} \quad (41)$$

and

$$\sigma_\varphi^2(r) = \sigma_\vartheta^2(r) = \frac{1}{6-q}(1+r^2)^{-\frac{1}{2}} \left(1 - \frac{q}{2}\frac{r^2}{1+r^2}\right), \quad (42)$$

so $\sigma_s^2(r) = \sigma_r^2(r) + \sigma_\varphi^2(r) + \sigma_\vartheta^2(r)$, then:

$$T = \pi \int_0^{+\infty} \rho_p(r_p) \sigma_i^2(r_p) r_p \, dr_p$$
$$= 2\pi \int_0^{+\infty} \rho(r) \sigma_b^2(r) r^2 \, dr = \frac{3\pi}{2^6}. \quad (43)$$

The kinetic energy is independent of $q$, by virtue of the virial theorem.

Radial cluster types have their greatest contribution to the kinetic energy at the center, considering the previous remark. Since the total kinetic energy is independent of $q$, tangential cluster types must have their prominent contribution at the edge. This feature is affirmed in the plots of the velocity dispersions in Fig. 5.

Moreover, a simple linear relation exists between the three velocity dispersions, namely

$$\sigma_p^2(r_p) = \sigma_t^2(r_p) + \frac{q}{12-q}\sigma_c^2(r_p). \quad (44)$$

This relation is independent of the projected distance $r_p$, and gives again a distinction between both cluster types. For radial clusters the line of sight velocity dispersion is always greater than the transverse velocity dispersion varying from equality if $q = 0$ until $\sigma_p^2 = \sigma_t^2 + \sigma_c^2/5$ if $q = 2$. On the contrary, in tangential clusters the transverse velocity dispersion is always greater than the radial velocity dispersion, now varying from equality in the isotropic case until $\sigma_t^2 = \sigma_p^2 + \sigma_c^2$ for $q \to -\infty$.

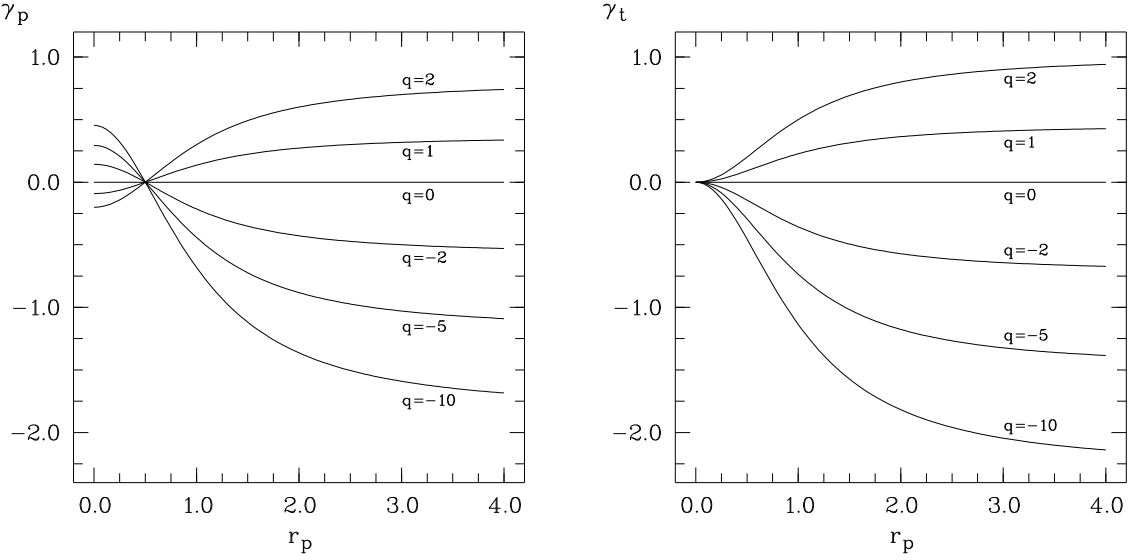

**Fig. 6.** The dependence of $\gamma_p$ and $\gamma_t$ as a function of $r_p$ for various $q$ values.

When we compare the projected velocity dispersions with the spatial velocity dispersions, given in Eq.(41) and Eq.(42), we notice the similarity between $\sigma_c^2(r_p)$ and $\sigma_r^2(r)$ on the one hand, and the dispersions $\sigma_t^2(r_p)$ or $\sigma_p^2(r_p)$ and $\sigma_\varphi^2(r_p)$ or $\sigma_\vartheta^2(r_p)$ on the other hand. So we define the parameters $\gamma_t$ and $\gamma_p$ analogous to $\beta$ (cfr. Eq. (6)):

$$\gamma_t = 1 - \frac{\sigma_t^2(r_p)}{\sigma_c^2(r_p)} = \frac{q}{12-q} \frac{5 r_p^2}{1+r_p^2} = (-1 + \frac{12}{12-q}) \frac{5 r_p^2}{1+r_p^2}, \quad (45)$$

$$\gamma_p = 1 - \frac{\sigma_p^2(r_p)}{\sigma_c^2(r_p)} = \frac{q}{12-q} \frac{4 r_p^2 - 1}{1+r_p^2} = (-1 + \frac{12}{12-q}) \frac{4 r_p^2 - 1}{1+r_p^2}. \quad (46)$$

In Fig.6 these relations are plotted for various $q$ values. The parameter $\gamma_t$ is a good diagnostic for the orbital structure: its sign is the same as the sign of $q$. In the center, however, it is zero for every value of $q$. The parameter $\gamma_p$ on the other hand is in the center region ($r_p < 0.5$) positive for negative $q$ values and negative for positive $q$ values, and is zero for $r_p = 0.5$. Therefore it is the parameter to use when only dispersions in the center are available. For $r_p \to \infty$ the parameter $\gamma_t$ varies from $-5$ for $q \to -\infty$ to 1 for $q = 2$ and the parameter $\gamma_p$ varies from $-4$ for $q \to -\infty$ to 0.8 for $q = 2$. So at the edge, $\gamma_t$ delivers us a better diagnostic since its range is wider. As usual, central velocity dispersions are less useful for determining the orbital structure than velocity dispersions obtained at the edge (in the unlikely case that both have the same accuracy). This is clearly visible in Fig.6, where the range in $\gamma_p$ and $\gamma_t$ is smaller in the center than at large radii.

From the relation (44) another peculiar result can be deduced, namely

$$\gamma_p = \gamma_t - \frac{q}{12-q} \quad (47)$$

which means that the left picture of Fig.6 can be obtained from the right one by pushing the curves up for negative $q$ and down for positive $q$.

## 5. The curtosis

In the same way as for the velocity dispersions, the fourth order moments $\tau_c^4(r_p)$, $\tau_t^4(r_p)$ of the marginal distributions can be derived. They are:

$$\tau_c^4(r_p) = \frac{168 - 22q + q^2}{2 \cdot 5 \cdot 7 \cdot (7-q)(6-q)(1+r_p^2)}, \quad (48)$$

$$\tau_t^4(r_p) = \frac{168 - 22q + q^2 - 4q(34-q)x^2 + 16q(q+2)x^4}{2 \cdot 5 \cdot 7 \cdot (7-q)(6-q)(1+r_p^2)}. \quad (49)$$

whereby $x^2 = r_p^2/(1+r_p^2)$. Instead of the fourth order moments, one preferably considers the dimensionless curtoses $\mathrm{cur}_c = \tau_c^4(r_p)/\sigma_c^4(r_p)$ and $\mathrm{cur}_t = \tau_t^4(r_p)/\sigma_t^4(r_p)$, which are a measure of the degree to which the marginal distributions are peaked: the greater the curtosis, the more peaked the distribution is. Usually this value is compared with the Gaussian distribution for which the curtosis is 3.

A remarkable feature is that the curtosis of the marginal distribution in the center direction is independent of the projected distance $r_p$. Figure 7 shows us the dependence of this curtosis on $q$. The curtosis is in every point smaller than three, so less than the normal distribution. Also radial clusters are a bit more peaked than tangential ones.

Figure 8 shows us the curtosis for the marginal distribution in the transversal direction. At every projected distance $r_p$, radial clusters have a greater curtosis than the tangential ones, as can also be noticed in the lower panels of Fig. 4. This was not the case for the curtosis of the line profile, which gave us in the center region a totally different picture, namely greater values are obtained for the tangential clusters! Just like the case of the velocity dispersions the scope at higher projected distances is greater for the marginal transversal distribution than for the line profile. This means again that from the proper motion one can get more accurate information.

Another peculiar feature are the very high values of the curtosis in the transversal direction for very radial systems. In the limit for $r_p \to \infty$ or for $x \to 1$ the transversal curtosis is $2^{11}(6-q)(4-q)/(\pi^2 \cdot 3 \cdot 5 \cdot 9(2-q)(7-q))$, which means that for $q \to 2$ this expression tends to infinity. So the two dimensional distribution becomes more and more one dimensional along the center axis, and has a very high top value. This has been checked by calculating the distribution (by numerical integration) for $r_p = 100$ and $q = 1.9$. The shape of the distribution for large $r_p$ and $q$ close to 2 is another confirmation of the earlier mentioned bad behaviour of the series approximation for these $q$ values.

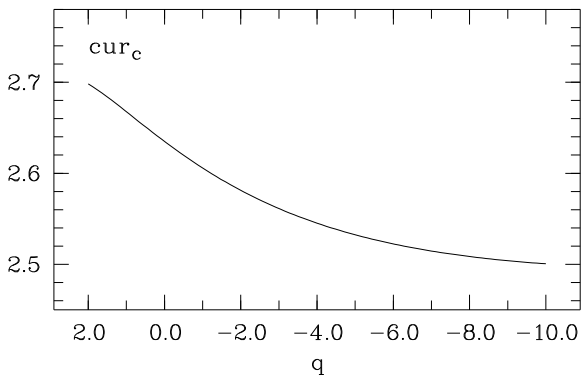

**Fig. 7.** The curtosis of the marginal distribution in the center direction as a function of $q$.

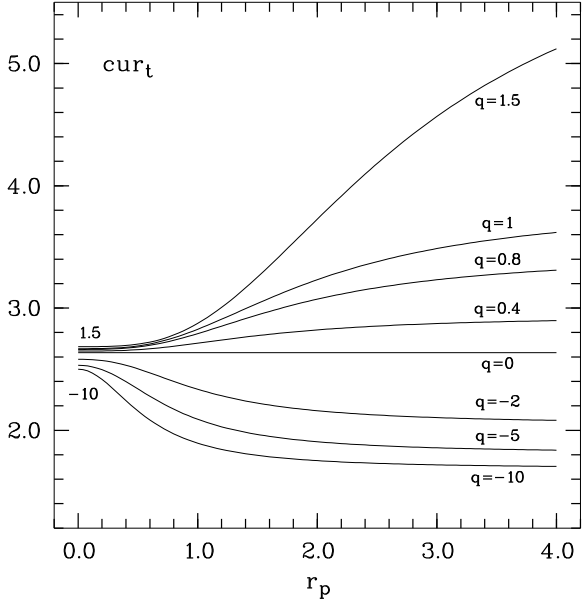

**Fig. 8.** The curtosis of the marginal distribution in the transversal direction as a function of the projected distance. Different values for $q$ are indicated.

## 6. The determination of the distance to a Plummer sphere

The proper motion dispersions (to the center of the system and perpendicular to it) depend on the distance to the system, on the projected distance from the center and on the orbital structure ($q$). The line of sight velocity dispersion however is independent of the distance, but depends also on the projected distance from the center and on the orbital structure. Combining these three velocity dispersions, it is in principle possible to determine the distance.

For the Plummer sphere, the distance can be obtained very easily as follows. Let $\mu_c$ and $\mu_t$ be the proper motion components as defined in sec. 2.2. Since $\gamma_t$ and $\gamma_p$ are defined as a function of projected velocity dispersions, they are observable at each distance $r_p$. Moreover $\gamma_t$ is independent of the distance to the system $\Delta$, so the relation (45) provides us with a $q$ for each $r_p$. The values of $q$ obtained this way, will be independent of $r_p$ only in the Plummer case. With this value of $q$, $\gamma_p$ can be calculated, and since

$$\gamma_p = 1 - \frac{\sigma_p^2}{\sigma_c^2} = 1 - \frac{\sigma_p^2}{\Delta^2 \mu_c^2}, \qquad (50)$$

with $\sigma_p$ known and independent of $\Delta$, the distance $\Delta$ can be determined.

The relative error $\delta\Delta/\Delta$ can be found by differentiating the first of the Eqs. (45) and (46) towards the dispersions and the second towards the parameter $q$. The result is

$$\frac{\delta\Delta}{\Delta} = \frac{\delta\sigma_p}{\sigma_p} + |\frac{\gamma_p}{\gamma_t}\frac{1-\gamma_t}{1-\gamma_p}|\frac{\delta\mu_t}{\mu_t} + |\frac{\gamma_p - \gamma_t}{\gamma_t(1-\gamma_p)}|\frac{\delta\mu_c}{\mu_c}, \qquad (51)$$

or expressed in measured quantities:

$$\begin{aligned}\frac{\delta\Delta}{\Delta} &= \frac{\delta\sigma_p}{\sigma_p} + |\frac{4r_p^2-1}{5r_p^2}\frac{\Delta^2\mu_t^2}{\sigma_p^2}|\frac{\delta\mu_t}{\mu_t} \\ &+ |-1+\frac{4r_p^2-1}{5r_p^2}\frac{\Delta^2\mu_t^2}{\sigma_p^2}|\frac{\delta\mu_c}{\mu_c},\end{aligned} \qquad (52)$$

which is formally independent of $q$.

Equation (51) will not hold when $\gamma_t = 0$. This occurs either when $r_p = 0$ or $q = 0$. In the former case (the center of the system), no appropriate measurements can be made. Even Eq. (52) shows this effect. The function of $r_p$ in this expression varies from 0.6 for $r_p = 1$ until 0.8 for $r_p$ going to infinity. The greatest influence on the relative error is then due to the factor $\Delta^2\mu_t^2/\sigma_p^2$, which makes it very clear that the larger the distance $\Delta$, the greater the error on the distance. In the latter case (isotropic), the three dispersions are all equal to:

$$\sigma_c^2 = \sigma_t^2 = \sigma_p^2 = \frac{3\pi}{64\sqrt{1+r_p^2}}. \qquad (53)$$

The distance $\Delta$ can then be obtained from $\Delta^2\mu_c^2 = \sigma_p^2$ or $\Delta^2\mu_t^2 = \sigma_p^2$ and has the relative error:

$$\frac{\delta\Delta}{\Delta} = \frac{\delta\sigma_p}{\sigma_p} + \frac{\delta\mu_c}{\mu_c} \qquad \text{or} \qquad \frac{\delta\Delta}{\Delta} = \frac{\delta\sigma_p}{\sigma_p} + \frac{\delta\mu_t}{\mu_t}. \qquad (54)$$

This formula comes in handy for a quick order-of-magnitude calculation of the relative error on the distance determination if the observational errors are known and the modelling errors are not too large.

## 7. Conclusions

In this paper we have explored the relation between the internal orbital structure of a spherical star cluster and the observed proper motion distributions. This we did for a particular family carry over, at least qualitatively, to most real clusters.

We find that simple inspection of the contour lines of these two-dimensional distributions suffices to get valuable information on the orbital structure, as is obvious from Figs. 2 and 3: the shape of the contour lines gives a direct and clear picture of the orbital structure. Moreover, we can get this information without the knowledge of the distance.

We also introduce two observable parameters, $\gamma_t$ and $\gamma_p$. One of them, $\gamma_t$, is the observational equivalent of Binney's anisotropy parameter $\beta$, and measures directly the amount of anisotropy. This is the quantitative equivalent of the statement in the previous paragraph.

As a by-product, we show that the moments of the proper motion distributions can be used to calculate these distributions. This means that it may suffice to calculate only the moments, and be reasonably confident that the proper motion distributions are meaningfully constrained. This will be necessary for modelling real clusters, where we can make no direct use of the specific formulae given here.

Last but not least, we show how to determine our distance to the system, if all three projected dispersions are known. This can be done, because the proper motion components are distance dependent and the line profile is distance independent, whereby they are all three interconnected through the orbital structure. The relative error on the distance depends on the relative error of the velocity dispersion of the line profile, and on the relative error of the two measured proper motion components which are multiplied by an appropiate function of the distance.

In a future paper, we will apply the concepts developed here to model globular clusters for which both proper motions and radial velocities are available.

*Acknowledgements.* We would like to thank the anonymous referee for a careful reading of the manuscript.

## 8. Appendices

### A. The value at the top

In this case, the value of the energy and angular momentum reduces to $E = \psi_p (\psi/\psi_p - \zeta^2)$ and $L = \sqrt{2\psi_p}\, r_p\, \zeta$. The Laplace-Mellin transform of the distribution function $F(E, L)$ does not depend on the velocity-component $\zeta$, and is:

$$\int_{-\infty}^{+\infty} F(E, L)\sqrt{2\psi_p}\, d\zeta$$
$$= \int_{-\infty}^{+\infty} \sqrt{2\psi_p}\, d\zeta\, \frac{1}{2\pi i} \int_{\alpha-i\infty}^{\alpha+i\infty} d\alpha$$
$$\times \frac{1}{2\pi i} \int_{\beta-i\infty}^{\beta+i\infty} d\beta\, \mathcal{LM}\{F\} e^{\alpha E} L^{-\beta}. \tag{A1}$$

So the integration towards $\zeta$ can be carried out:

$$\int_{-\infty}^{+\infty} \sqrt{2\psi_p}\, d\zeta\, e^{\alpha \psi_p(\frac{\psi}{\psi_p} - \zeta^2)} (\sqrt{2\psi_p}\, r_p\, \zeta)^{-\beta}$$
$$= e^{\alpha \psi}\, r_p^{-\beta}\, (\frac{\alpha}{2})^{\frac{\beta-1}{2}}\, \Gamma(\frac{1-\beta}{2}). \tag{A2}$$

If the density $\rho(\psi, r)$, which is a function of the potential $\psi$ and the distance $r$, can be expressed as a product of the form $\rho(\psi, r) = \psi^p\, g(r)$ then the Laplace-Mellin transform of $F(E, L)$ can be written as (see Dejonghe, 1986):

$$\mathcal{LM}\{F\} = \frac{\Gamma(p+1)}{(2\pi)^{\frac{3}{2}}}\, \frac{\alpha^{\frac{1}{2}(1-\beta)-p}}{\Gamma(1-\frac{1}{2}\beta)}\, 2^{\frac{\beta}{2}}\, \mathcal{M}\{g(r)\}. \tag{A3}$$

In the case of the Plummer model this transform is

$$\mathcal{LM}\{F\} = \frac{3\,\Gamma(6-q)}{4\pi\,(2\pi)^{\frac{3}{2}}}\, \frac{\alpha^{\frac{1}{2}(1-\beta)-5+q}}{\Gamma(1-\frac{\beta}{2})}\, 2^{\frac{\beta}{2}}\, \mathcal{M}\{g(r)\}, \tag{A4}$$

and together with the Mellin transform we have

$$\mathcal{M}\{g(r)\} = \int_0^{+\infty} r^{\beta-1}(1+r^2)^{-\frac{q}{2}}\, dr = \frac{\Gamma(\frac{1}{2}\beta)\Gamma(\frac{1}{2}(q-\beta))}{2\,\Gamma(\frac{1}{2}q)}. \tag{A5}$$

Equation (A1) then becomes:

$$\int_{-\infty}^{+\infty} F(E, L)\sqrt{2\psi_p}\, d\zeta$$
$$= \frac{3\Gamma(6-q)\sqrt{2}}{4(2\pi)^{\frac{5}{2}}\Gamma(\frac{1}{2}q)}\, \frac{1}{2\pi i} \int_{-i\infty}^{+i\infty} d\beta \frac{\Gamma(\frac{1-\beta}{2})\Gamma(\frac{\beta}{2})\Gamma(\frac{q-\beta}{2})}{\Gamma(1-\frac{\beta}{2})}\, r_p^{-\beta}$$
$$\times \frac{1}{2\pi i} \int_{-i\infty}^{+i\infty} e^{\alpha \psi}\, \alpha^{q-5}\, d\alpha. \tag{A6}$$

Since the last integral is nothing else than an inverse Laplace transform $\mathcal{L}^{-1}(\alpha^{q-5}) = \psi^{4-q}/\Gamma(5-q)$ and the first one is Barnes' integral (Slater, 1966, p22) Eq. (A6) results in:

$$\int_{-\infty}^{+\infty} F(E, L)\sqrt{2\psi_p}\, d\zeta$$
$$= \frac{3(5-q)}{8\pi^2}\, \psi^{4-q}\, {}_2F_1\left(\frac{q}{2}, \frac{1}{2}; 1; -r_p^2\right). \tag{A7}$$

To calculate the PM-distribution at the top we still need to perform an integration through the cluster, which is

$$\int_{-\infty}^{+\infty} \psi^{4-q}\, dz = \int_{r_p}^{r_{\max}} \frac{\psi^{4-q}}{\sqrt{(r^2 - r_p^2)}}\, dr^2$$
$$= \sqrt{\pi}\, \psi_p^{3-q}\, \frac{\Gamma(\frac{3}{2} - \frac{q}{2})}{\Gamma(2 - \frac{q}{2})} \tag{A8}$$

so that Eq. (A7) leads to

$$\mathrm{pm}_{r_p}(0, 0) = \frac{3(5-q)}{4\sqrt{\pi}}\, \frac{\Gamma(\frac{3}{2} - \frac{q}{2})}{\Gamma(2 - \frac{q}{2})}\, \psi_p^{-q}\, {}_2F_1\left(\frac{q}{2}, \frac{1}{2}; 1; -r_p^2\right). \tag{A9}$$

With the relation (Exton, 1978, p17)

$${}_2F_1(a, b; c; z) = (1-z)^{-a}\, {}_2F_1(a, c-b; c; -\frac{z}{1-z}) \tag{A10}$$

the above expression can be transformed into

$$\mathrm{pm}_{r_p}(0, 0) = \frac{3(5-q)}{4\sqrt{\pi}}\, \frac{\Gamma(\frac{3}{2} - \frac{q}{2})}{\Gamma(2 - \frac{q}{2})}\, {}_2F_1(\frac{q}{2}, \frac{1}{2}; 1; \frac{r_p^2}{1+r_p^2}) \tag{A11}$$

which gives us the value at the top as a simple hypergeometric function, with argument $r_p^2/(1+r_p^2)$.

## B. The moments of the distribution of the proper motion

The moments of the distribution function $F(E, L)$ are defined as:

$$\mu_{n,m,l} = M \iiint F(E, L)\, v_r^n\, v_\varphi^m\, v_\vartheta^l\, dv_r\, dv_\varphi\, dv_\vartheta \tag{B1}$$

with $M$ the mass, $E$ the energy and $L$ the absolute value of the angular momentum. The integration is carried out over all the possible values of the velocities. All odd moments are zero because the total system has no net streaming. The anisotropic moments can be defined as:

$$\mu_{2n,2j} = 2\pi M \iint F(E, L)\, v_r^{2n}\, v_T^{2j+1}\, dv_r\, dv_T \tag{B2}$$

with $v_r$ the velocity component pointed towards the center of the system and $v_T$ the tangential velocity: $v_T = (v_\varphi^2 + v_\vartheta^2)^{1/2}$. The relation between both defined moments can easily be shown by putting Eq. (B1) in cylindrical coordinates:

$$\mu_{2n,2m,2l} = \frac{1}{\pi}\, B\!\left(m + \frac{1}{2}, l + \frac{1}{2}\right) \mu_{2n,2(m+l)} \tag{B3}$$

The anisotropic moments can be expressed in function of the potential and the distance $r$ of the center through (Dejonghe, 1986):

$$\mu_{2n,2m}(\psi, r)$$
$$= \frac{2^{m+n}}{\sqrt{\pi}} \frac{\Gamma(n+\frac{1}{2})}{\Gamma(m+n)}$$
$$\times \int_0^\psi (\psi - \psi')^{m+n-1}\, D_{r^2}^m (r^{2m} \mu_{0,0}(\psi', r)))\, d\psi' \tag{B4}$$

with $m + n \geq 1$ and with $D_{r^2}^m$ representing the $m$th derivative towards $r^2$.

We wish to give an expression of the moments of the distribution function of the proper motion. So we have to transform the velocity components to the coordinate system as defined in section 2.2. We have:

$$v_c^p v_t^s = (a v_r + b v_\varphi)^p v_\vartheta^s = \sum_{i=0}^p \binom{p}{i} a^{p-i}\, b^i\, v_r^{p-i}\, v_\varphi^i\, v_\vartheta^s \tag{B5}$$

with $a = r_p/r$ and $b = z/r$. Taking the mean value of this expression, considering that all odd moments have to be zero, so the indices $p, i$ and $s$ need to be even, we have:

$$\langle v_c^{2p} v_t^{2s} \rangle = \sum_{i=0}^p \binom{2p}{2i} a^{2(p-i)}\, b^{2i}\, \mu_{2(p-i),2i,2s} \tag{B6}$$

With the use of Eqs. (B3) and (B4), this becomes:

$$\langle v_c^{2p} v_t^{2s} \rangle$$
$$= \frac{1}{\pi} \sum_{i=0}^p \binom{p}{i} a^{2(p-i)}\, b^{2i}\, \frac{\Gamma(s+\frac{1}{2})\Gamma(p+\frac{1}{2}) 2^{p+s}}{\Gamma(p+s)\Gamma(s+i++1)}$$
$$\times \int_0^\psi (\psi + \psi')^{p+s-1}\, D_{r^2}^{s+i}(r^{2(s+i)} \mu_{0,0}(\psi', r))\, d\psi'. \tag{B7}$$

By means of the integral formula of Cauchy we can transform the $(s+i)$th derivative into a contour integral around $r^2$, so that the summation according to $i$ can be calculated, yielding:

$$\langle v_c^{2p} v_t^{2s} \rangle$$
$$= \frac{\Gamma(s+\frac{1}{2})\Gamma(p+\frac{1}{2}) 2^{p+s}}{\pi\,\Gamma(p+s)} \frac{1}{2\pi i} \int_{\mathcal{C}(r^2)} \frac{dt}{(t-r^2)^{s+1}}$$
$$\times \int_0^\psi (\psi - \psi')^{p+s-1}\, t^s \left(\frac{t-r_p^2}{t-r^2}\right)^p \mu_{0,0}(\psi', t)\, d\psi' \tag{B8}$$

This is a general expression, valuable for all kind of $\mu_{0,0}(\psi, t)$.

The expression of the mass density in the Plummer model is given by $\mu_{0,0}(\psi', t) = (3/4\pi)\, \psi'^{5-q}(1+t)^{-q/2}$, so that the integration towards $\psi'$ in Eq. (B4) is a simple beta-function, namely: $(\psi(r))^{p+s+6-q}\, B(6-q, p+s)$. Let the result of the integration through the cluster be represented by $\langle v_c v_t \rangle^{2p,2s}$, or:

$$\langle v_c v_t \rangle^{2p,2s} = \int_{r_p}^{+\infty} \frac{dr^2}{\sqrt{r^2 - r_p^2}} \langle v_c^{2p} v_t^{2s} \rangle. \tag{B9}$$

To calculate this integration, we consider new variables $v$ and $u$ by means of $v = (t-r_p^2)/(1+r_p^2)$ and $u = (r^2-r_p^2)/(1+r_p^2)$, and another quite obvious notation for the product of gamma functions (Slater, 1966, p41), so that Eq. (B9) leads to:

$$\langle v_c v_t \rangle^{2p,2s} = \frac{3 \cdot 2^{p+s}}{4\pi^2} \Gamma\!\left(\begin{array}{c} p+\frac{1}{2}, s+\frac{1}{2}, 6-q \\ p+s+6-q \end{array}\right)(1+r_p^2)^{-\frac{p+s}{2}-2}$$
$$\times \frac{1}{2\pi i} \int_0^{+\infty} u^{-\frac{1}{2}} (1+u)^{-\frac{1}{2}(p+s+5-q)}\, du$$
$$\times \int_{\mathcal{C}(u)} \frac{(v+x^2)^s\, v^p\, (1+v)^{-\frac{q}{2}}}{(v-u)^{p+s+1}}\, dv \tag{B10}$$

with $x^2 = r_p^2/(1+r_p^2)$, which is nothing else than the cumulative mass density. The special case $p = s = 0$ yields $\langle v_c v_t \rangle^{0,0} = 1/(\pi(1+r_p^2)^2)$ which is nothing else than the projected mass density at the location $r_p$.

It is necessary to normalise the moments towards the maximum velocity and towards the projected mass density. Its notation will be $\langle \chi\, \eta \rangle^{2p,2s}$ and with the aid of Eqs. (9) and (11) this normalisation is defined as

$$\langle \chi\, \eta \rangle^{2p,2s} = \frac{\pi}{2^{p+s}} (1+r_p^2)^{\frac{1}{2}(p+s)+2} \langle v_c v_t \rangle^{2p,2s} \tag{B11}$$

so that $\langle \chi\, \eta \rangle^{0,0} = 1$. To calculate the integral in Eq. (B10) the expression $(v+x^2)^s$ is expanded in $\sum_{i=0}^s \binom{s}{i} x^{2i} v^{s-i}$ and the contour integral is changed back into the derivatives, yielding:

$$\langle \chi\, \eta \rangle^{2p,2s} = \frac{3}{4\pi} \Gamma\!\left(\begin{array}{c} p+\frac{1}{2}, s+\frac{1}{2}, 6-q \\ p+s+6-q, p+s+1 \end{array}\right) \sum_{i=0}^s \binom{s}{i} x^{2i}$$
$$\times \sum_{j=i}^{p+s} \binom{p+s}{j} \left(\frac{q}{2}\right)_j \frac{(-1)^j \Gamma(p+s-i+1)}{\Gamma(j-i+1)}$$
$$\times \int_0^{+\infty} u^{j-i-\frac{1}{2}} (1+u)^{-\frac{1}{2}(p+s+5)-j}\, du \tag{B12}$$

distribution of the proper motion are given by a finite double summation. By changing the summation index $j$ to $k$ with $k = j - i$ and so $m = p + s - i$, the last summation is a combination of gamma functions and a finite hypergeometric series with three parameters in the numerator and two in the denominator. The above expression of the moments becomes:

$$\langle \chi\,\eta \rangle^{2p,2s} = \frac{3}{4\sqrt{\pi}} \Gamma \binom{p+\frac{1}{2},\,s+\frac{1}{2},\,6-q}{p+s+6-q}$$
$$\times \sum_{i=0}^{s} \binom{s}{i} x^{2i} \frac{(-1)^i \Gamma(\frac{1}{2}(p+s)+i+2)\,(\frac{q}{2})_i}{\Gamma(i+1)\Gamma(\frac{1}{2}(p+s+5)+i)}$$
$$\times \ {}_3F_2 \binom{\frac{q}{2}+i,\,\frac{1}{2},\,-(p+s-i);}{i+1,\,\frac{1}{2}(p+s+5)+i;} \ 1 \ \Bigg). \tag{B13}$$

This is in fact a double summation, one with argument $x^2$ and one with argument 1. This expression can be transformed into a Kampé de Fériet function by considering the following two properties:

– the relation which transforms a ${}_{A+1}F_{B+1}$ hypergeometric function into an integration of a ${}_A F_B$ hypergeometric function is:

$${}_{A+1}F_{B+1}\,(c,(a);d,(b);z)$$
$$= \frac{\Gamma(d)}{\Gamma(c)\Gamma(d-c)}$$
$$\times \int_0^1 t^{c-1}\,(1-t)^{d-c-1}\,{}_A F_B\,((a);(b);tz)\,dt \tag{B14}$$

with $(a) = a_1, a_2, \ldots, a_A$ the different factors in the numerator and $(b) = b_1, b_2, \ldots, b_B$ the different factors in the denominator.

– a finite ${}_3F_2$ series with argument 1 can be transformed into another ${}_3F_2$ series with argument 1 after multiplying the series with some appropriate gamma functions (Slater, 1966, Chap.4).

If we choose out of all the other possible series, that one for which the expressions $c$ and $d$ in Eq. (B14) do not contain the summation index $i$, then it is possible to change the order of integration and summation in the expression of $\langle \chi\,\eta \rangle^{2p,2s}$. This leads after some calculations to the final expression of the moments of the distribution of the proper motion:

$$\langle \chi\,\eta \rangle^{2p,2s}$$
$$= \Gamma \binom{6-q,\,p+\frac{1}{2},\,s+\frac{1}{2},\,p+s+\frac{1}{2},\,\frac{1}{2}(p+s)+2,\,\frac{3}{2}(p+s)-\frac{q}{2}+3}{p+s+6-q,\,p+s+1,\,\frac{1}{2}(p+s-q)+3,\,\frac{3}{2}(p+s)+\frac{5}{2}}$$
$$\times \ \frac{3}{4\pi}\,\mathbf{F}^{1:1,2}_{1:0,1}\binom{\frac{1}{2}(p+s)+2\,:\,-s\,;\,\frac{1}{2},\,-(p+s)\,;}{-(p+s)+\frac{1}{2}\,:\,-\,;\,\frac{1}{2}(p+s-q)+3\,;}\,x^2,1\,\Bigg)(\text{B15})$$

with $\mathbf{F}$ a Kampé de Fériet function (Exton, 1978, p24). This is in fact a generalization of the hypergeometric function: it is a double summation with argument $x^2$ concerning the first summation index $i$ and argument 1 concerning the second summation index $j$. After the symbol F, the numbers of parameters in the numerator (upper ones) and in the denominator (lower ones) are indicated. The first numbers concern the summation index $i + j$, and the corresponding parameters are first mentioned. They are separated by a colon from the other two groups. The first group concerns the first summation index $i$ (here one parameter in the numerator, which is $-s$, so this is a finite summation) and zero parameters in the denominator. The second concerns the second summation index $j$, here with two parameters in the numerator, which are $1/2$ and $-(p+s)$ (so again a finite summation), and one in the denominator, which is $(p+s-q)/2+3$. The separation of these parameters is now indicated by a semi-colon. Special cases of the Kampé de Fériet functions are the four better known Appel functions.

### C. Outer regions of the PM-distribution

We will put forward an approximation of the PM-distribution for velocities close to the escape velocities, this means for the outer regions of the PM-distribution. Here is $v^2 = \chi^2 + \eta^2 \to 1$ so that $\zeta^2$ is very small. Since there is a double prescription of $F(E,L)$, given in Eqs. (4) and (5), two kinds of approaches are needed.

#### C.1. Small values of $r_p$

In the case of small values of $r_p$ the greatest contributions to the outer regions of the PM-distributions come from the stars with a radial velocity close to the escape velocity. For those stars the angular momentum vector is small and consequently $L^2 << E$, and we can restrict the hypergeometric function in Eq.(4) to the first term:

$$F(E,L) \approx \frac{3\,\Gamma(6-q)}{2(2\pi)^{\frac{5}{2}}\,\Gamma(\frac{9}{2}-q)}\,E^{\frac{7}{2}-q} \tag{C1}$$

By using the usual symbols and normalisations, the PM-distribution becomes:

$$\text{pm}_{r_p}(\chi,\eta)$$
$$= \frac{2\pi}{\psi_p^3}\frac{3\,\Gamma(6-q)}{2(2\pi)^{\frac{5}{2}}\,\Gamma(\frac{9}{2}-q)} \int_{r_p}^{r_{max}} \frac{dr^2}{\sqrt{r^2-r_p^2}}$$
$$\times \int_{-\zeta_{max}}^{\zeta_{max}} (\psi_p)^{\frac{7}{2}-q}\,(\zeta_{max}^2-\zeta^2)^{\frac{7}{2}-q}\,\sqrt{2\psi_p}\,d\zeta. \tag{C2}$$

The integration towards $\zeta$ is a beta-function, so that the above expression becomes:

$$\text{pm}_{r_p}(\chi,\eta)$$
$$= \frac{3(5-q)}{4\pi}\,\psi_p^{1-q} \int_{r_p}^{r_{max}} \frac{dr^2}{\sqrt{r^2-r_p^2}}\,\left(\frac{\psi}{\psi_p}-v^2\right)^{4-q}. \tag{C3}$$

Transforming the $r$-variable into $\psi$ considering that $1+r_{max}^2 = 1/\psi_{max}^2 = 1/(\psi_p^2\,v^4)$, leads to:

$$\text{pm}_{r_p}(\chi,\eta)$$
$$= \frac{3(5-q)}{4\pi}\,\psi_p^{1-q} \int_{\psi_p\,v^2}^{\psi_p} \frac{d\psi}{\psi^3} \frac{\psi\,\psi_p}{\sqrt{\psi_p^2-\psi^2}}\,\left(\frac{\psi}{\psi_p}-v^2\right)^{4-q}. \tag{C4}$$

Integrating from 0 to 1 with the substitution $\psi = \psi_p(v^2+u(1-v^2))$ results in:

$$\text{pm}_{r_p}(\chi,\eta)$$
$$= \frac{3(5-q)}{2\pi}\,\psi_p^{-q}$$
$$\times \int_0^1 \frac{(1-v^2)^{\frac{9}{2}-q}\,u^{4-q}\,du}{(v^2+u(1-v^2))^2\,(1-u)^{1/2}\,((1-v^2)u+1+v^2)^{1/2}}. \tag{C5}$$

so that the first factor in the denominator goes to 1 and the last one goes to $\sqrt{2}$. The integration towards $u$ is a beta-function, namely $B(\frac{1}{2}, 5-q)$ so that we finally have:

$$\text{pm}_{r_p}(\chi, \eta) = \frac{3\,\Gamma(6-q)\,\psi_p^{-q}}{2\sqrt{2\pi}\,\Gamma(\frac{11}{2}-q)}\,(1-v^2)^{\frac{9}{2}-q}. \tag{C6}$$

### C.2. Large values of $r_p$

In this case the most important contributions to the outer regions of the PM-distribution come from those stars whose angular momentum is great. So here is $L^2 \gg E$ and so we restrict the distribution function in Eq. (5) to the first term of the hypergeometric function:

$$F(E, L) = \frac{3\,\Gamma(6-q)\,2^{\frac{q}{2}}}{2\,(2\pi)^{\frac{5}{2}}\Gamma(\frac{9}{2}-\frac{q}{2})\Gamma(1-\frac{q}{2})} E^{\frac{7-q}{2}} L^{-q} \tag{C7}$$

whereby

$$L^{-q} \approx (2\psi_p)^{-q/2}\,r_p^{-q}\,(\eta^2 + \zeta^2)^{-q/2} \tag{C8}$$

since in these regions the $z$-values are very small with regard to $r_p$. Also $v^2 = \chi^2 + \eta^2$ is nearly 1. This means that $\zeta^2$ is very small. If $\eta$ is small, then the angular momentum is small and we are back in the previous case. If $\eta$ is large, we can then approximate $L$ by:

$$L^{-q} \approx (2\psi_p)^{-q/2}\,r_p^{-q}\,(\eta^2)^{-q/2} \tag{C9}$$

so that the PM-distribution becomes:

$$\text{pm}_{r_p}(\chi, \eta)$$
$$= \frac{2\pi}{\psi_p^3} \frac{3\,\Gamma(6-q)\,r_p^{-q}\,(\eta^2)^{-\frac{q}{2}}}{2\,(2\pi)^{\frac{5}{2}}\Gamma(\frac{9}{2}-\frac{q}{2})\Gamma(1-\frac{q}{2})}(\psi_p)^{\frac{7}{2}-q}$$
$$\times \int_{r_p}^{r_{max}} \frac{dr^2}{\sqrt{r^2-r_p^2}} \int_{-\zeta_{max}}^{\zeta_{max}} (\zeta_{max}^2 - \zeta^2)^{\frac{7-q}{2}}\,\sqrt{2\psi_p}\,d\zeta. \tag{C10}$$

Integration towards $\zeta$ is a simple beta-function, yielding:

$$\text{pm}_{r_p}(\chi, \eta)$$
$$= \frac{3\,\Gamma(6-q)\,r_p^{-q}\,(\eta^2)^{-\frac{q}{2}}\,\psi_p^{1-q}}{4\pi\,\Gamma(5-\frac{q}{2})\Gamma(1-\frac{q}{2})}$$
$$\times \int_{r_p}^{r_{max}} \frac{dr^2}{\sqrt{r^2-r_p^2}}\left(\frac{\psi}{\psi_p} - v^2\right)^{4-\frac{q}{2}}. \tag{C11}$$

Transforming the $r$-variable into $\psi$ considering that $1+r_{max}^2 = 1/\psi_p^2 v^4$ and integrating from 0 to 1 with the substitution $\psi = \psi_p(v^2 + u(1-v^2))$ results in:

$$\text{pm}_{r_p}(\chi, \eta)$$
$$= \frac{3\,\Gamma(6-q)\,r_p^{-q}\,(\eta^2)^{-\frac{q}{2}}\,\psi_p^{-q}}{2\pi\,\Gamma(5-\frac{q}{2})\Gamma(1-\frac{q}{2})}$$
$$\times \int_0^1 \frac{(1-v^2)^{\frac{9}{2}-\frac{q}{2}}\,u^{4-\frac{q}{2}}\,du}{(v^2+u(1-v^2))^2\,(1-u)^{1/2}\,((1-v^2)u+1+v^2)^{1/2}}. \tag{C12}$$

For $v^2 \to 1$ this leads to:

$$\text{pm}_{r_p}(\chi, \eta)$$
$$= \frac{3\Gamma(6-q)}{2\sqrt{2\pi}\,\Gamma(\frac{11}{2}-\frac{q}{2})}\left(\frac{r_p^2}{1+r_p^2}\right)^{-\frac{q}{2}} \eta^{-\frac{q}{2}}\,(1-v^2)^{\frac{9}{2}-\frac{q}{2}}. \tag{C13}$$